\documentclass[10pt]{iopart}
\begin{document}
\title{ Phase transitions in simple and not so simple binary fluids}
\author{A Parola$^{\dagger,\ast}$, D Pini$^{\ddagger,\ast}$, 
L Reatto$^{\ddagger,\ast}$ and M Tau$^{\parallel,\ast}$}
\address {
$^\dagger$ Dipartimento di Scienze, Universit\`a dell'Insubria, Como Italy\\
$^\ddagger$ Dipartimento di Fisica, Universit\`a di Milano, Milano, Italy \\
$^\parallel$ Dipartimento di Fisica, Universit\`a di Parma, Parma, Italy \\
$^\ast$ Istituto Nazionale per la Fisica della Materia, Genova, Italy}
\begin{abstract}
Compared to pure fluids, binary mixtures display a very diverse phase
behavior, which depends sensitively on the parameters of 
the microscopic potential. 
Here we investigate the phase diagrams of simple model mixtures 
by use of a microscopic implementation of the renormalization
group technique. First, we consider a symmetric mixture with attractive 
interactions, possibly relevant for describing fluids of
molecules with internal degrees of freedom. 
Despite the simplicity of the model, slightly tuning the strength 
of the interactions between unlike species drastically changes the topology 
of the phase boundary, forcing or inhibiting demixing, and brings about 
several interesting features such as 
double critical points, tricritical points,
and coexistence domains enclosing ``islands'' of homogeneous, mixed fluid.  
Homogeneous phase separation in mixtures can be driven also 
by purely repulsive interactions. As an example, we consider 
a model of soft particles  
which has been adopted to describe binary polymer solutions. This is shown 
to display demixing (fluid-fluid) transition at sufficiently high density. 
The nature and the physical properties of the corresponding phase transition
are investigated. 
\end{abstract}

The phase diagram of binary mixtures may show several transition 
lines, usually related to the physically different processes of demixing and
liquid-vapor phase separation\cite{row}. 
While the ordinary liquid-vapor phase transition
is driven by the presence of attractive interactions, demixing may be
occur also in purely repulsive fluids and even in strongly asymmetric, 
athermal (i.e. hard-core) systems. However, inspecting the 
composition of the phases at coexistence generally reveals a more structured
scenario where both density and composition of the phase-separated fluids
differ. Even in rare-gas mixtures, like the Ne-Kr system, 
the character of the transition can change from liquid-vapor at low pressure
to mainly mixing-demixing at high pressure. This behavior 
requires a generalization of the concept of {\sl order parameter}, allowing
for linear combinations of density and concentration fluctuations:
By moving along a critical line the nature of the order parameter,
i.e. the weights of density and concentration fluctuations in the
linear combination, smoothly changes signaling the change in the
physical character of the transition. 
As a result, an ample variety of phase diagram topologies 
are possible in mixtures, and it is not surprising that small changes 
in the interaction parameters may have a crucial effect in 
determining important physical properties of the system, like the
miscibility of two fluids at given pressure conditions. 

In ordinary simple fluids, this sensitivity of the phase diagram to
small changes in the interaction has not attracted much interest
in the past, because we have little possibility to affect 
the form of the interaction at the molecular level. 
However, a considerably richer scenario opens up in the
framework of complex fluids. If the constituents of our 
mixtures are polymers, colloids or micelles, their mutual interaction is
mediated by the solvent and then it depends on several properties
which may be suitably modified, like polymer-chain lengths  
or salt concentration. It is therefore important to investigate
how (small) changes in the interparticle potentials affect 
the global phase diagram and the structure of the mixture
and whether it is possible to switch from one topology of the phase
diagram to another
by simply acting on the form of the effective interactions:
physically this means that by small changes in the properties 
of the solvent we may ``turn on'' strong concentration
fluctuations in the system, driving mixing or demixing. 

Liquid-state theory offers several methods which allow the
investigation of this problem\cite{book}. A classical approach, which
has been widely exploited in the past, is to resort to 
mean field theory (MFT) which has the considerable advantage to 
require very modest (if any) computational effort and to 
provide the full phase diagram of the model. However, the
main drawbacks of MFT are the complete neglect of fluctuations,
extremely relevant close to phase transitions,
and the insensitivity to the {\it shape} of the interactions,
which are known to play an important role in complex fluids
\cite{note}. A more sophisticated route to the theoretical investigation
of phase diagrams is represented by the integral equation approach.
This method is designed to study correlation functions and by now
the best integral equations (the modified hypernetted chain or
the Rogers-Young equations) provide a remarkably accurate 
representation of the structure of the model. 
Thermodynamics instead is obtained via the use of sum rules, but
the resulting phase diagram often depends on the chosen route
and suffers from a lack of convergence of the theory in the critical region,
casting doubts on the accuracy of the method. The most reliable
method to investigate phase coexistence in fluids is probably
the Gibbs ensemble Monte Carlo method which proved quite 
successful in several different situations. However, simulations  
are rather time consuming, mainly because they 
provide information on a single thermodynamic state of the system.
The possibility of mapping a large portion of the phase diagram is
therefore hindered by the computational burden.

In this work, we present an investigation of the phase diagram of 
few simple models of binary mixtures by use of the Hierarchical Reference
Theory of Fluids (HRT)\cite{adv}, 
a method explicitly devised for studying the thermodynamics
of liquids and gases which allows for the development of long-range 
fluctuations and consistently enforces the convexity of the free energy.
The method is based on a splitting of the interaction 
in a repulsive-core part, treated as a reference system, 
and a ``long-range'' tail 
which is supposed to drive possible phase transitions. The effects 
on the thermodynamics of the long-range part are then introduced
in the free energy recursively, starting from short wavelengths
in a way which resembles the renormalization group strategy.
HRT is especially useful to determine the phase boundaries and the
equation of state of microscopic models where energy driven phase transitions
are at play. It has been tested both in fluid and lattice systems,
and accurate results have been obtained. A generalization to mixtures
has been already numerically implemented both for lattice and off-lattice
systems\cite{hrt}. Here we discuss the phase diagram of a symmetric 
mixture when the ratio $\Delta$ between
the unlike and the like attractive interactions, is varied. 
The interparticle interactions $w_{ij}(r)$ were modelled by a hard-core Yukawa
(HCY) potential such that $w_{ij}(r)=+\infty$ for $r<\sigma$ 
and $w_{ij}(r)=-\epsilon_{ij}\exp[-z(r-\sigma)]/r$ for $r>\sigma$, where
$i$, $j$ refer to the particle species, $\epsilon_{11}=\epsilon_{22}$, 
$\epsilon_{12}=\Delta\epsilon_{11}$, $\sigma$ is the particle diameter,
and $z$ is the inverse-range parameter, which we fixed at $z=1.8\sigma^{-1}$.
Full details regarding
the computations will be soon available\cite{new}.
A similar model has been already studied by Monte Carlo methods\cite{wilding}.
This system offers an interesting example of competition between liquid-vapor
and mixing-demixing transitions.
As it can be appreciated from Fig. \ref{figsymm}, an increase by just
10\% of the unlike interaction $\Delta$ leads to the formation of
a stable mixed phase at liquid density and equimolar concentration.
While for $\Delta=0.6$ the two fluids are miscible only in the gas
phase, for $\Delta=0.67$, at the same temperature, the two fluids
mix also in the liquid region, i.e. at density up to
$\rho\sigma^3\sim0.5$.
In such a hypothetical system, by slightly tuning the interaction we may
therefore inhibit demixing at high pressures. 
Criticality in this symmetric binary mixture presents a rich variety 
of different behaviors.
For instance, at $\Delta=0.67$, 
up to six critical points are present for each temperature in 
a small range below $T^*=1.03$. Well 
above this characteristic temperature the system 
displays an ordinary consolute critical point which then turns to 
a tricritical point (an artifact due to the symmetry of the model).
Below $T^*$ one pair of (liquid-vapor) critical points merges, giving rise to 
a double critical point and then disappear. The remaining four
critical points then merge in pairs (at $T^*\sim 1.0235$) 
leaving an ``island" in the density-concentration plane where the
mixed liquid is stable. 
On further lowering the temperature, this one-phase domain shrinks, until it 
is eventually swallowed by the coexistence region. The phase diagram in
the $(\rho,T)$ plane at equimolar concentration is shown in Fig. \ref{figrhot}.
 
Up to now we analyzed mixtures where  phase transitions are driven
by attractive potentials. However, colloidal suspensions are often 
characterized by purely repulsive interactions\cite{louis} and we may inquire 
whether also in this case 
the topology of the phase diagram depends on the details of the interaction.
Some sensitivity on the specific form of the 
interaction in these systems has been already conjectured regarding 
the freezing transition\cite{loewen1}. We have investigated, by means 
of HRT, the gaussian-core model (GCM), whose interparticle potential 
has the form:
\begin{equation}
\beta v(r)=\epsilon \exp{(-r^2/\sigma^2)}
\end{equation}
which has been proposed as
an effective model for dilute and semi-dilute polymer solutions\cite{hansen1}.
Here the interaction is purely repulsive and has entropic origin,
being due to the self-avoidance of the polymer coils. 
The strength of the interaction depends on the number of
monomers in the polymer. A characteristic feature of this
model is the allowance of full overlap between ``particles''. This
apparently unphysical feature originates from the fact that in the
effective model considered, a particle just represents the center
of mass of the polymer, i.e. a geometric point and not a physical particle. 
No phase transition between homogeneous phases is observed 
in this model\cite{loewen2}. 
A two-component system of gaussian-core ``molecules''
has also been investigated both when the unlike interaction
$\epsilon_{12}$ is larger than the like ones $\epsilon_{11}=\epsilon_{22}$
and in the opposite case\cite{finken,evans}. Note that 
in this model with purely repulsive interactions
the liquid vapor transition does not occur, and  
in order to trigger the demixing transition  positive non additivity in the
core radii combination rule would be required if
$\Delta\equiv \epsilon_{12}/\epsilon_{11}<1$. 

HRT appears particularly suitable for treating this class of effective 
interactions because it has 
been shown\cite{hansen2,loewen2} that the momentum dependence of the
partial structure factors is well represented by the simple 
mean field approximation in a large portion of the phase diagram. 
This is of great help in devising the closure relation necessary
to write the formally exact HRT equation in a manageable form. 
In Fig.\ref{figgaus} (left panel) 
we report the density-concentration phase diagram of the 
GCM for few choices of the interaction parameters, compared with the 
results obtained by MFT and HNC equation\cite{finken}. 
We remark that, for given $\Delta$, the MFT phase boundaries for different
interaction strengths $\epsilon_{11}$ collapse into the same curve when
plotted in the units used in the figure\cite{finken}, while this is not 
the case with 
either HNC or HRT. However, even according to these theories 
the changes induced by a significant increase in the interaction
are just quantitative, while the overall topology of the phase diagram 
remains unaltered.  When the strength of the unlike interaction grows,
HRT pushes the phase boundary to higher densities with respect to 
HNC. A remarkable flattening of the phase boundaries is also predicted
by HRT.
Actually, the HRT binodals reported in the figure show portions   
that are completely straight, but this is due to the finite density mesh
used in the computation. This applies also to the small jumps in the form 
of the binodals. 
 
Accurate Monte Carlo simulations
on this model may be useful to resolve the discrepancy between the
two approaches. A snapshot of correlations is reported in the left panel
where some clustering of the minority species,
due to the softness of the repulsive interaction,
can be seen close to the demixing transition\cite{finken}.

The effective hamiltonian approach to complex fluids is intimately
related to the possibility of tracing out degrees of freedom.
However, we point out that in the class of models characterized by
soft interactions this powerful technique, which is
at the very basis of the concept of depletion interactions, 
may lead to unphysical results. As an example, we consider the
gaussian-core mixture where an accurate free energy 
functional has been already introduced\cite{loewen2,hansen2,evans}. If we ``freeze''
the degrees of freedom of the $N_1$ particles of species `1' 
in a given, arbitrary, configuration $\{{\bf R}_i\}$, the free energy functional 
of the gas of particles `2' in this inhomogeneous external potential
can be accurately represented in mean-field approximation:
\begin{eqnarray}
F[\rho({\bf r})]=F_{id}[\rho({\bf r})]&+&{1\over 2} \int d{\bf r}_1 \int d{\bf r}_2 \rho
({\bf r}_1) \rho({\bf r}_2)v_{22}({\bf r}_1-{\bf r}_2) \nonumber\\
&+&\sum_i \int d{\bf r} v_{12}({\bf r}-{\bf R}_i)\rho({\bf r})
\label{energy}
\end{eqnarray}
This functional must be minimized in order to obtain the actual
density profile $\rho({\bf r})$ of particles `2' subject to the 
number conservation constraint $\int d{\bf r} \rho({\bf r})=N_2$.
The calculation can be carried out analytically for weak inhomogeneity
and the resulting density profile is given by:
\begin{eqnarray}
\rho({\bf r})&=&\rho_0\left [ 1- \sum_i \Delta\rho({\bf r}-{\bf R}_i)\right ]
\nonumber\\
\Delta\rho({\bf r})&=&\int {d{\bf q}\over (2\pi)^3} \exp(i{\bf q}\cdot{\bf r})
{\beta v_{12}(q)\over 1+\rho_0\beta v_{22}(q)}
\label{density}
\end{eqnarray}
where $\rho_0$ is determined by imposing the constraint.
By substituting Eq.~(\ref{density}) into Eq.~(\ref{energy}) we obtain 
the two-body {\it effective} interaction experienced by particles `1'
due to the presence of particles `2', which turns out to be globally attractive:
\begin{equation}
v_{\rm eff}(q)=-\beta\rho_2 {v_{12}(q)^2\over  1+\rho_2\beta v_{22}(q)}
\label{effect}
\end{equation}
In the limit $v_{22}(r)=0$ and weak unlike interaction
the procedure becomes exact
and still predicts purely attractive interactions, which may drive the system
to a collapse: a clearly unphysical result for a model with
only {\it repulsive} terms ! Actually,  
many-body contributions to the effective interaction, which  are
obtained by including higher orders in the minimization of 
the free energy but are usually disregarded, eventually 
stabilize the effective model. 
The analysis of this toy model suggests that 
many-particle terms in the effective hamiltonian are likely to play a crucial
role in systems where particle overlap is 
not inhibited.
This is also supported by the analysis brought forth in Ref.\cite{evans2}. 
In that work, effective two-body forces in ternary gaussian mixtures were 
investigated by considering two big particles in a solvent of smaller 
particles at different compositions. Strongly attractive sovent-mediated
interactions were found both in the special case of a pure solvent, and 
in the more general case where the solvent is itself a mixture of two species,
when a dramatic increase in the effective interaction can occurr 
as a consequence of preferential adsorbtion by the big particles. 
We observe that the result for $v_{\rm eff}$ reported in Eq.~(4)
of\cite{evans2} for a pure solvent 
and very small solvent/solute size ratio coincides with that obtained
in the same limit by Eq.~(\ref{effect}) of the present work.
  Due to the above mentioned sensitivity
of the phase diagram of mixtures on the form of the interactions, 
we conclude that many-body forces should be taken into account
in deriving the effective hamiltonians, especially in soft core systems. 

In summary, we have shown how a classic textbook subject like
the thermodynamics of binary mixtures may present 
still open problems when applied to the case of complex fluids. 
In particular, attention has been focused on the sensitivity of
the phase diagram to the strength of the unlike interaction and
to the possibility of shaping the phase boundaries. Very soft 
effective potentials, like the GCM,  are ideal systems 
for understanding the mechanisms underlying phase 
segregation in purely repulsive, athermal fluids.

\Figures

\begin{figure}
\caption{Density-concentration coexistence regions of the symmetric HCY
mixture for: $T=1.02$, $\Delta=0.6$ (left panel) and $T=1.02$,
$\Delta=0.67$ (right panel). The quantities $\rho$ and $x$ are respectively
the total density $\rho=\rho_{1}+\rho_{2}$ in reduced units and the relative
concentration $x=\rho_{2}/\rho$. The gray scale is a measure of pressure
(black: low pressure, white high pressure). Few tie lines are also shown.
}
\label{figsymm}
\end{figure}

\begin{figure}
\caption{Equimolar section of the density-temperature phase diagram of
the symmetric HCY mixture for $\Delta=0.67$.
}
\label{figrhot}
\end{figure}

\begin{figure}
\caption{Left panel:
Phase boundaries of the symmetric gaussian core mixture in the 
density-concentration plane for $\epsilon_{11}=0.1$ (triangles), 
$\epsilon_{11}=1$ (squares), $\epsilon_{11}=2$ (circles)
and $\epsilon_{12}=2\epsilon_{11}$. 
Full symbols: HRT results; open symbols: HNC data; solid line: 
mean-field binodal.  Here $\rho=\rho_{1}+\rho_{2}$, $x=\rho_{2}/\rho$, and
$\tilde v_{11}(0)=\pi^{3/2}\epsilon_{11}\sigma^3$.
Right panel:
Two-body partial correlation functions of the same model in the case 
$\epsilon_{11}=0.1$ obtained via HRT (lines) 
for two thermodynamic states with $x=0.9$ and different densities:
$\rho\sigma^{3}=2$ (lower curves) and $\rho\sigma^{3}=4$ (upper curves). 
Dots represent HNC data for $\rho\sigma^3=2$.}
\label{figgaus}
\end{figure}


\begin{thebibliography}{99}
\bibitem{note} A typical example are depletion (effective) interactions,
characterized by a remarkably short range which makes MFT 
grossly inaccurate.
\end{thebibliography}
\end{document}